\documentclass[a4paper,11pt]{article}
\pdfoutput=1 
\usepackage{jcappub} 
\usepackage[T1]{fontenc} 
\usepackage{amssymb,amsmath,amsfonts}
\usepackage{hyperref}  
\usepackage{float}
\usepackage{graphicx}
\usepackage{xcolor}
\usepackage{tensor}
\usepackage[utf8]{inputenc}
\usepackage{caption,subfig}
\usepackage{feynmf}
\usepackage{tikz}
\usepackage{nicefrac}

\title{One-loop renormalization in Galileon effective field theory}
\author[a]{Lavinia Heisenberg,}
\author[b]{Christian F. Steinwachs}
\affiliation[a]{Institute for Theoretical Physics, 
	ETH Zurich, Wolfgang-Pauli-Strasse 27, 8093, Zurich, Switzerland}
\affiliation[b]{Physikalisches Institut, Albert-Ludwigs-Universit\"at Freiburg,
	Hermann-Herder-Str.~3, 79104 Freiburg, Germany}
\emailAdd{lavinia.heisenberg@phys.ethz.ch}
\emailAdd{christian.steinwachs@physik.uni-freiburg.de}

\abstract{
We investigate the renormalization structure of scalar Galileons in flat spacetime. We explicitly calculate the ultraviolet divergent one-loop contributions to the $2$-point, $3$-point, $4$-point, and $5$-point functions. We discuss the structure of the counterterms and their hierarchy within an effective field theory expansion. We comment on different resummation schemes, including a geometric resummation for which our results could be generalized to arbitrary $n$-point functions. 
}

\begin{document}
	\allowdisplaybreaks[1]
	\maketitle
	\flushbottom
	%

\section{Introduction}
There are many modified theories of gravity, which invoke new dynamical degrees of freedom. Among the cosmologically most relevant approaches, which introduce an additional propagating scalar degree of freedom, are $f(R)$ gravity and scalar-tensor theories, see e.g. \cite{Sotiriou2010,DeFelice2010,Clifton2012,Nojiri2017}. The renormalization structure of these models on a general curved background has been derived in \cite{Barvinsky1993,Shapiro1995,Steinwachs2011,Steinwachs2012,Rham2013,Kamenshchik2015,Ruf2018,Ruf2018c}. Besides scalar field models, generalized vector field models have also been investigated in the cosmological context \cite{Ford1989,Golovnev2008,Esposito-Farese2010,Tasinato2014,Heisenberg2014,BeltranJimenez2016,Allys2016,Heisenberg2018,Heisenberg2018a}. The renormalization structure of the generalized Proca theory in curved spacetime has been discussed in \cite{Buchbinder2007,Toms2014,Toms2015,Belokogne2016, Buchbinder2017,Ruf2018b,Ruf2018a}.

In this article we analyze the renormalization structure of the scalar Galileon model \cite{Dvali2000,Nicolis2009,Deffayet2009a,Rham2011}. Quantum aspects of the Galileon have been studied previously in \cite{Nicolis2004,Hinterbichler2010,PaulaNetto2012,Rham2013,Heisenberg2014a,Brouzakis2014,Brouzakis2014a,Kampf2014,Pirtskhalava2015}.
From an effective field theory point of view, the Galileon correspond to the most general Lagrangian for a scalar field $\pi$, which gives rise to second order equations of motion and is invariant under the Galilean transformation
\begin{align}
\pi\to\pi+c+v_{\mu}x^{\mu},\label{Galsym}
\end{align}
with a constant $c$ and a constant vector $v_{\mu}$.
The second-order nature of the equations of motion ensures the absence of propagating ghost degrees of freedom.
In four spacetime dimensions, only five non-trivial tree-level operator structures of the Galileon scalar field are possible.
The Galileon symmetry \eqref{Galsym} prevents the generation of loop induced operators with less than two derivatives per field and, in particular, no renormalization of the tree-level operators is required. Nevertheless, from the point of view of an effective field theory expansion, the counterterm structure of the Galileon effective field theory is very interesting.

We derive the off-shell one-loop divergences for the Galileon in flat spacetime within the traditional momentum space Feynman diagrammatic approach.
Since the derivation of the one-loop counterterms in the $\overline{\mathrm{MS}}$ scheme only requires the extraction of the ultraviolet divergent (UV) part of the $1$PI one-loop diagrams, we make use of the recently proposed combinatorial algorithm, specifically designed for an efficient evaluation of one-loop UV divergences in higher derivative theories \cite{Steinwachs2019}. We discuss the hierarchy of loop induced operators in an effective field theory expansion and comment on various resummation schemes. 
Other interesting aspects connected to the effective field theory of higher derivative scalar fields and the Galileon were analyzed in \cite{Kampf2013,Kampf2014,Cheung2015,Cachazo2015,Cheung2015a,Cheung2016,Cheung2017,Carrasco2019}.
Within the bootstrap program, on-shell recursion relations are use to calculate scattering amplitudes and the requirements of analyticity, Lorentz invariance and a particular behavior of the tree-level scattering amplitude IR singularities in the soft an colinear limits completely characterize the theory.
In this way, even without having to specify any Lagrangian or spell out any symmetry principle, the landscape of scalar EFT's can be scanned and classified \cite{Cheung2015,Cheung2015a,Cheung2017}.

This article is organized as follows: In Sec.~\ref{SecAct}, we introduce the Galileon model in flat spacetime. In Sec. \ref{SecFeynman} we derive the momentum space Feynman rules. In Sec.~\ref{SecOneLoop}, we classify the topologies of all one-loop $1$PI diagrams contributing to the UV divergent parts of the $1$-point, $2$-point, $3$-point, $4$-point, and $5$-point functions. We briefly review our method for the explicit calculation of the off-shell one-loop divergences. Due to the high number of terms in the higher $n$-point functions, we provide the actual results in a separate ancillary file.\footnote{The ancillary file (.txt format) can be accessed via this \href{https://arxiv.org/src/1909.04662v1/anc/AncillaryFile_OneLoop_OffShell_UVDiv_Galileon.txt}{link}.}
In Sec.~\ref{SecRepResCheck} we discuss the representation of the result and perform several crosschecks.
In Sec.~\ref{SecRenEff}, we analyze the renormalization structure of the scalar Galileon and discuss various resummation schemes. Finally, in Sec.~\ref{SecCon}, we summarize our main results and give a brief outlook on further developments.  

\section{Galileon action}
\label{SecAct}
The action functional for the scalar  Galileon field $\pi(x)$ in $d=4$ flat Euclidean space with metric $\delta_{\mu\nu}=\text{diag}(1,1,1,1)$ reads,
\begin{align}
S[\pi]={}&\int\mathop{}\!\mathrm{d}^{4}x\mathcal{L}(\pi,\partial\pi,\partial^2\pi),\qquad \mathcal{L}=\sum_{i=1}^5\mathcal{L}_{i},\label{Gact}\\
\mathcal{L}_{1}={}&c_1 M^3\pi,\label{L0}\\
\mathcal{L}_{2}={}&c_2\pi\epsilon^{\mu\nu\rho\sigma}\tensor{\epsilon}{^{\alpha}_{\nu\rho\sigma}}\pi_{\mu\alpha},\label{L1}\\
\mathcal{L}_{3}={}&\frac{c_3}{M^3}\pi\epsilon^{\mu\nu\rho\sigma}\tensor{\epsilon}{^{\alpha\beta}_{\rho\sigma}}\pi_{\mu\alpha}\pi_{\nu\beta},\\
\mathcal{L}_{4}={}&\frac{c_4}{M^6}\pi\epsilon^{\mu\nu\rho\sigma}\tensor{\epsilon}{^{\alpha\beta\gamma}_{\sigma}}\pi_{\mu\alpha}\pi_{\nu\beta}\pi_{\rho\gamma},\\
\mathcal{L}_{5}={}&\frac{c_5}{M^9}\pi\epsilon^{\mu\nu\rho\sigma}\tensor{\epsilon}{^{\alpha\beta\gamma\delta}}\pi_{\mu\alpha}\pi_{\nu\beta}\pi_{\rho\gamma}\pi_{\sigma\delta}\label{L4}\,.
\end{align}
The symmetric tensor $\pi_{\mu\nu}$ results from second derivatives of $\pi(x)$ and is defined as
\begin{align}
\pi_{\mu\nu}:=\partial_{\mu}\partial_{\nu}\pi.
\end{align}
Denoting by $[O]$ the mass dimension of an object $O$ in units $\hbar=c=1$, we have $[M]=[\pi]=1$ $[\mathcal{L}_{i}]=4$ and $[S]=[c_{i}]=0$.
We neglect tadpole contributions and choose $c_2=1/12$ such that $\pi$ has a canonically normalized propagator.\footnote{This is no restriction since the dimensionless coupling constant $c_2$ can always be absorbed by a redefinition of the Galileon field $\pi\to(c_2)^{1/2}\pi$ and a corresponding rescaling of the coupling constants $c_i\to c_i(c_2)^{-i/2}$.} 

\section{Momentum space Feynman rules}
\label{SecFeynman}
We define the scalar contraction between two momenta $k_{i}^{\mu}$ and $k_{j}^{\mu}$ as
\begin{align}
(k_{i}\cdot k_{j}):={}&k_{i}^{\mu}k_{j}^{\nu}\eta_{\mu\nu}. 
\end{align}
For the contraction of two identical momenta $i=j$, we define the square as
\begin{align}
k_{i}^2:={}&(k_{i}\cdot k_{i}).\label{momsqr}
\end{align}
The massless propagator in momentum space reads
\begin{align}
P(k):=\frac{1}{k^2},\label{PropP}
\end{align}
with its graphical representation 
\begin{center}
\begin{tikzpicture}
\node at (0,0) {$P(k)=$};
\begin{scope}[xshift=0.8cm]
\draw[black](0,0)--(1,0);
\node at (1.1,0) {$.$};
\end{scope}
\end{tikzpicture}
\end{center}

\noindent The momentum space vertices are defined with all momenta incoming 
\begin{align}
V^{(3)}(k_1,k_2,k_3)={}&\frac{c_3}{M^3}\tensor{\varepsilon}{_{\mu\nu\rho\sigma}}\tensor{\varepsilon}{_{\alpha\beta}^{\rho\sigma}}k_{1}^{\mu}k_{1}^{\alpha}k_{2}^{\nu}k_{2}^{\beta}+\text{cyclic}(1,2,3)\nonumber\\
={}&2\frac{c_3}{M^3}\mathrm{G}(k_1,k_2)+\text{cyclic}(1,2,3)\label{V3},\\
V^{(4)}(k_1,k_2,k_3,k_4)={}&\frac{c_4}{M^6}\tensor{\varepsilon}{_{\mu\nu\rho\sigma}}\tensor{\varepsilon}{_{\alpha\beta\gamma}^{\sigma}}k_{1}^{\mu}k_{1}^{\alpha}k_{2}^{\nu}k_{2}^{\beta}k_{3}^{\rho}k_{3}^{\gamma}+\text{cyclic}(1,2,3,4),\nonumber\\
={}&\frac{c_4}{M^6}\mathrm{G}(k_1,k_2,k_3)+\text{cyclic}(1,2,3,4),\\
V^{(5)}(k_1,k_2,k_3,k_4,k_5)={}&\frac{c_5}{M^9}\tensor{\varepsilon}{_{\mu\nu\rho\sigma}}\tensor{\varepsilon}{_{\alpha\beta\gamma\delta}}k_{1}^{\mu}k_{1}^{\alpha}k_{2}^{\nu}k_{2}^{\beta}k_{3}^{\rho}k_{3}^{\gamma}k_{4}^{\sigma}k_{4}^{\delta}+\text{cyclic}(1,2,3,4,5)\nonumber\\
={}&\frac{c_5}{M^9}\mathrm{G}(k_1,k_2,k_3,k_4)+\text{cyclic}(1,2,3,4,5)
,\label{V5}
\end{align}
and are graphically represented as

\begin{center}
\begin{tikzpicture}
\node at (0,0) {$V^{(3)}=$};
\begin{scope}[xshift=1.5cm]
\filldraw[black] (0:0) circle (0.7mm);
\draw[black] (60:0) -- (60:0.8);
\draw[black] (180:0) -- (180:0.8);
\draw[black] (300:0) -- (300:0.8);
\node at (0.7,0) {$,$};
\end{scope}

\node at (4,0) {$V^{(4)}=$};
\begin{scope}[xshift=5cm]
\filldraw[black] (0:0) circle (0.7mm);
\draw[black] (45:0) -- (45:0.8);
\draw[black] (135:0) -- (135:0.8);
\draw[black] (225:0) -- (225:0.8);
\draw[black] (315:0) -- (315:0.8);
\node at (0.7,0) {$,$};
\end{scope}

\node at (8,0) {$V^{(5)}=$};
\begin{scope}[xshift=9cm]
\filldraw[black] (0:0) circle (0.7mm);
\draw[black] (72:0) -- (72:0.8);
\draw[black] (72+72:0) -- (72+72:0.8);
\draw[black] (72+72+72:0) -- (72+72+72:0.8);
\draw[black] (72+72+72+72:0) -- (72+72+72+72:0.8);
\draw[black] (72+72+72+72+72:0) -- (72+72+72+72+72:0.8);
\node at (1,0) {$.$};
\end{scope}
\end{tikzpicture}
\end{center}
\noindent Here $G(k_1,\ldots, k_n)$ denotes the determinant of the $n\times n$ Gram matrix
\begin{align}
G_{ij}:=\left(k_i\cdot k_j\right),\quad i,j,=1,\ldots,n.\label{Gram}
\end{align}

\section{One-loop calculation}
\label{SecOneLoop}
\subsection{Topologies of one-loop $1$PI Feynman diagrams}
The topologies of the one-loop $1$PI Feynman diagrams with up to five external legs directly follow from the Feynman rules \eqref{PropP}-\eqref{V5}. The two numbers $(S,M)$ below each diagram denote the symmetry factor $S$ and the multiplicity $M$ of the corresponding diagram, respectively.

\subsubsection*{Two-point function}
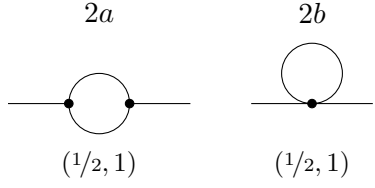
\begin{center}
\begin{tikzpicture}[scale=0.8]
\begin{scope}
	\draw[black](0,0)--(1,0);
	\node[black] (2a) at (1.5,1.5) {$2a$ };
	\node[black] (2a) at (1.5,-1) {\small $(\nicefrac{1}{2},1)$};
	\filldraw[black] (1,0) circle (0.7mm);
	\draw[black] (1.5,0) circle (0.5cm);
	\filldraw[black] (2,0) circle (0.7mm);
	\draw[black](2,0)--(3,0);
\end{scope}
	
\begin{scope}[xshift=5cm]
	\draw[black](-1,0)--(1,0);
	\node[black] (2b) at (0,1.5) {$2b$ };
	\node[black] (2b) at (0,-1) {\small $(\nicefrac{1}{2},1)$};
	\filldraw[black] (0,0) circle (0.7mm);
	\draw[black] (0,0.5) circle (0.5cm);
\end{scope}
\end{tikzpicture}
\end{center}
	
\subsubsection*{Three-point function}
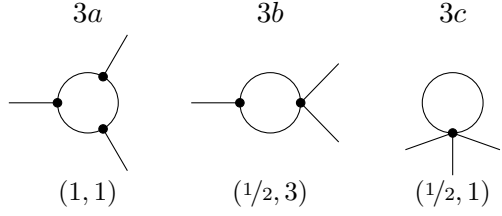
\begin{center}
\begin{tikzpicture}[scale=0.8]
\begin{scope}
\draw[black] (0,0) circle (0.5cm);
\node[black] (3a) at (0,1.5) {$3a$ };
\node[black] (3a) at (0,-1.5) {\small $(1,1)$};
\filldraw[black] (60:0.5) circle (0.7mm);
\draw[black] (60:0.5) -- (60:1.3);
\filldraw[black] (180:0.5) circle (0.7mm);
\draw[black] (180:0.5) -- (180:1.3);
\filldraw[black] (300:0.5) circle (0.7mm);
\draw[black] (300:0.5) -- (300:1.3);
\end{scope}

\begin{scope}[xshift=3cm]
\draw[black] (0,0) circle (0.5cm);
\node[black] (3b) at (0,1.5) {$3b$ };
\node[black] (3b) at (0,-1.5) {\small $(\nicefrac{1}{2},3)$};
\filldraw[black] (180:0.5) circle (0.7mm);
\draw[black] (180:0.5) -- (180:1.3);
\filldraw[black] (0:0.5) circle (0.7mm);
\draw[black] (0:0.5) -- (30:1.3);
\draw[black] (0:0.5) -- (-30:1.3);
\end{scope}

\begin{scope}[xshift=6cm]
\draw[black] (0,0) circle (0.5cm);
\node[black] (3c) at (0,1.5) {$3c$ };
\node[black] (3c) at (0,-1.5) {\small $(\nicefrac{1}{2},1)$};
\filldraw[black] (270:0.5) circle (0.7mm);
\draw[black] (270:0.5) -- (270:1.2);
\draw[black] (270:0.5) -- (225:1.1);
\draw[black] (270:0.5) -- (315:1.1);
\end{scope}
\end{tikzpicture}
\end{center}
	
\subsubsection*{Four-point function}
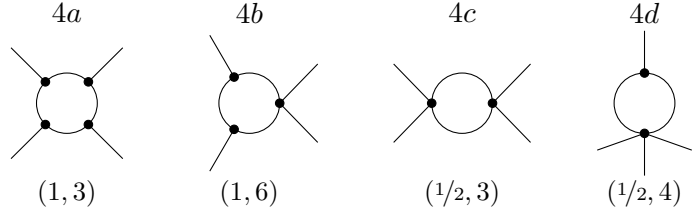
\begin{center}
\begin{tikzpicture}[scale=0.8]
\begin{scope}
\draw[black] (0,0) circle (0.5cm);
\node[black] (4a) at (0,1.5) {$4a$ };
\node[black] (4a) at (0,-1.5) {\small $(1,3)$};
\filldraw[black] (45:0.5) circle (0.7mm);
\draw[black] (45:0.5) -- (45:1.3);
\filldraw[black] (135:0.5) circle (0.7mm);
\draw[black] (135:0.5) -- (135:1.3);
\filldraw[black] (225:0.5) circle (0.7mm);
\draw[black] (225:0.5) -- (225:1.3);
\filldraw[black] (315:0.5) circle (0.7mm);
\draw[black] (315:0.5) -- (315:1.3);
\end{scope}

\begin{scope}[xshift=3cm]
\draw[black] (0,0) circle (0.5cm);
\node[black] (4b) at (0,1.5) {$4b$ };
\node[black] (4b) at (0,-1.5) {\small $(1,6)$};
\filldraw[black] (120:0.5) circle (0.7mm);
\draw[black] (120:0.5) -- (120:1.3);
\filldraw[black] (240:0.5) circle (0.7mm);
\draw[black] (240:0.5) -- (240:1.3);
\filldraw[black] (0:0.5) circle (0.7mm);
\draw[black] (0:0.5) -- (30:1.3);
\draw[black] (0:0.5) -- (-30:1.3);
\end{scope}

\begin{scope}[xshift=6.5cm]
\node[black] (4c) at (0,1.5) {$4c$ };
\node[black] (4c) at (0,-1.5) {\small $(\nicefrac{1}{2},3)$};
\draw[black] (0,0) circle (0.5cm);
\filldraw[black] (180:0.5) circle (0.7mm);
\draw[black] (180:0.5) -- (210:1.3);
\draw[black] (180:0.5) -- (150:1.3);
\filldraw[black] (0:0.5) circle (0.7mm);
\draw[black] (0:0.5) -- (30:1.3);
\draw[black] (0:0.5) -- (-30:1.3);
\end{scope}

\begin{scope}[xshift=9.5cm]
\node[black] (4d) at (0,1.5) {$4d$ };
\node[black] (4d) at (0,-1.5) {\small $(\nicefrac{1}{2},4)$};
\draw[black] (0,0) circle (0.5cm);
\filldraw[black] (270:0.5) circle (0.7mm);
\draw[black] (270:0.5) -- (270:1.2);
\draw[black] (270:0.5) -- (225:1.1);
\draw[black] (270:0.5) -- (315:1.1);
\filldraw[black] (90:0.5) circle (0.7mm);
\draw[black] (90:0.5) -- (90:1.2);
\end{scope}
\end{tikzpicture}
\end{center}

\subsubsection*{Five-point function}

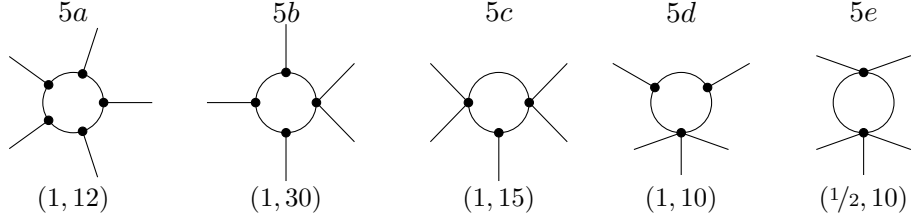
\begin{center}
\begin{tikzpicture}[scale=0.8]
\begin{scope}
\draw[black] (0,0) circle (0.5cm);
\node[black] (5a) at (0,1.5) {$5a$ };
\node[black] (5a) at (0,-1.6) {\small $(1,12)$};
\filldraw[black] (72:0.5) circle (0.7mm);
\draw[black] (72:0.5) -- (72:1.3);
\filldraw[black] (144:0.5) circle (0.7mm);
\draw[black] (144:0.5) -- (144:1.3);
\filldraw[black] (216:0.5) circle (0.7mm);
\draw[black] (216:0.5) -- (216:1.3);
\filldraw[black] (288:0.5) circle (0.7mm);
\draw[black] (288:0.5) -- (288:1.3);
\filldraw[black] (360:0.5) circle (0.7mm);
\draw[black] (360:0.5) -- (360:1.3);
\end{scope}

\begin{scope}[xshift=3.5cm]
\draw[black] (0,0) circle (0.5cm);
\node[black] (5b) at (0,1.5) {$5b$ };
\node[black] (5b) at (0,-1.6) {\small $(1,30)$};
\filldraw[black] (90:0.5) circle (0.7mm);
\draw[black] (90:0.5) -- (90:1.3);
\filldraw[black] (180:0.5) circle (0.7mm);
\draw[black] (180:0.5) -- (180:1.3);
\filldraw[black] (270:0.5) circle (0.7mm);
\draw[black] (270:0.5) -- (270:1.3);
\filldraw[black] (0:0.5) circle (0.7mm);
\draw[black] (0:0.5) -- (30:1.3);
\draw[black] (0:0.5) -- (-30:1.3);
\end{scope}

\begin{scope}[xshift=7cm]
\draw[black] (0,0) circle (0.5cm);
\node[black] (5c) at (0,1.5) {$5c$ };
\node[black] (5c) at (0,-1.6) {\small $(1,15)$};
\filldraw[black] (180:0.5) circle (0.7mm);
\draw[black] (180:0.5) -- (210:1.3);
\draw[black] (180:0.5) -- (150:1.3);
\filldraw[black] (270:0.5) circle (0.7mm);
\draw[black] (270:0.5) -- (270:1.3);
\filldraw[black] (0:0.5) circle (0.7mm);
\draw[black] (0:0.5) -- (30:1.3);
\draw[black] (0:0.5) -- (-30:1.3);
\end{scope}

\begin{scope}[xshift=10cm]
\draw[black] (0,0) circle (0.5cm);
\node[black] (5d) at (0,1.5) {$5d$ };
\node[black] (5d) at (0,-1.6) {\small $(1,10)$};
\filldraw[black] (270:0.5) circle (0.7mm);
\draw[black] (270:0.5) -- (270:1.2);
\draw[black] (270:0.5) -- (225:1.1);
\draw[black] (270:0.5) -- (315:1.1);
\filldraw[black] (150:0.5) circle (0.7mm);
\draw[black](150:0.5)--(150:1.3);
\filldraw[black] (30:0.5) circle (0.7mm);
\draw[black](30:0.5)--(30:1.3);
\end{scope}

\begin{scope}[xshift=13cm]
\draw[black] (0,0) circle (0.5cm);
\node[black] (5e) at (0,1.5) {$5e$ };
\node[black] (5e) at (0,-1.6) {\small $(\nicefrac{1}{2},10)$};
\filldraw[black] (270:0.5) circle (0.7mm);
\draw[black] (270:0.5) -- (270:1.2);
\draw[black] (270:0.5) -- (225:1.1);
\draw[black] (270:0.5) -- (315:1.1);
\filldraw[black] (90:0.5) circle (0.7mm);
\draw[black](90:0.5)--(45:1.1);
\draw[black](90:0.5)--(135:1.1);
\end{scope}
\end{tikzpicture}
\end{center}

\subsection{Extraction of one-loop UV divergences}
\label{algorithm}
Each of the $1$PI one-loop diagrams $X=2a,\ldots,5e$ correspond to a sum of scalar Feynman integrals $I_{X}=\sum_{k}I_{k}$, where the $I_{k}$ are of the form
\begin{align}
I_{k}:=K(Q)\int\frac{\mathrm{d}^d\ell}{(2\pi)^{d}}(\ell^2)^\lambda\prod_{i=1}^{n}\frac{\left(q_i\cdot \ell\right)^{\sigma_i}}{D_i^{\delta_i}},\label{FeynIntn}
\end{align}
with $Q:=\{q_i\}$, $i=1,\ldots,n$, the loop momentum $\ell^{\mu}$, the positive integers $\lambda$, $\sigma_i$ and $\delta_i$ and the inverse propagators,
\begin{align}
D_i:=\left(\ell+q_i\right)^2, \quad i=1,\ldots,n,
\end{align}
labeled by the combination $q_i^{\mu}$ of external momenta $k_{i}^{\mu}$, obeying momentum conservation,
\begin{align}
q_i^{\mu}:=\sum_{j=1}^{i}k_j^{\mu},\qquad q_{n}^{\mu}=q_0^{\mu}=\sum_{j=1}^{n}k_j ^{\mu}=0.\label{exq}
\end{align} 
Here, $K(Q)$ denotes a kinematic scalar independent of $\ell^{\mu}$, which for the massless theory \eqref{Gact} only depends on the external momenta $Q$ and can be written in the form
\begin{align}
K(Q)=c\,\prod_{1 \leq i\leq j\leq n}\left(q_i\cdot q_j\right)^{r_{ij}},
\end{align}  
with the $n(n+1)/2$ integer numbers $r_{ij}$ and a constant numerical coefficient $c$.
We use dimensional regularization in $d=4-2\varepsilon$ dimensions and isolate the ultraviolet divergent parts of the one-loop Feynman integrals \eqref{FeynIntn} as poles $1/\varepsilon$ in the limit $\varepsilon\to0$.
An efficient way to extract the one-loop ultraviolet divergences in higher derivative theories is based on an expansion of the propagators in the integrand of \eqref{FeynIntn} around vanishing external momenta and power counting arguments.
The propagator expanded around $q_i^{\mu}=0$ up to $\mathcal{O}\left(\ell^u\right)$ yields
\begin{align}
\frac{1}{D_i}=\sum_{0\leq 2\alpha+\beta+2\leq u}\left(
\begin{array}{c}
\alpha+\beta\\\alpha
\end{array}
\right)\frac{\left(-q_i^2\right)^\alpha\left(-2\ell\cdot q_i\right)^\beta}{\left(\ell^2+m_\mathrm{IR}^2\right)^{1+\alpha+\beta}}.\label{PropEx}
\end{align}
Note that we have introduced an auxiliary infrared regulating mass $m_{\mathrm{IR}}$ in \eqref{PropEx}. The final result for the ultraviolet divergences is independent of $m_{\mathrm{IR}}$.
By using \eqref{PropEx}, the integrals \eqref{FeynIntn} are reduce to a sum of vacuum integrals of the form
\begin{align}
I_{\mathrm{vac}}=&K(Q)\int\frac{\mathrm{d}^d\ell}{(2\pi)^{d}}\frac{(\ell^2)^\lambda\prod_{i=1}^{n}\left(q_i\cdot \ell\right)^{\sigma_i}}{\left(\ell^2+m_{\mathrm{IR}}^2\right)^\beta},\qquad 2\omega:=\sum_{i=1}^{n}\sigma_i.\label{LogDivInt}
\end{align}
In $d=4$, logarithmically divergent integrals have a superficial degree of divergence 
\begin{align}
\chi_{\mathrm{div}}:=4+2(\omega+\lambda-\beta)=0.
\end{align} 
In \cite{Steinwachs2019}, a combinatorial formula was derived by which the remaining logarithmically divergent integrals can be easily evaluated 
\begin{align}
I_{\mathrm{div}}^{\mathrm{vac}}
={}&\Lambda\sum_{k\in\mathcal{P}(2\omega;\sigma_{ij})}\frac{C_k}{P_{\omega}(d)}\prod_{1\leq i\leq j\leq n}(q_i\cdot q_j)^{\sigma_{ij}^{k}}.\label{DivIntFin}
\end{align}
Here, the sum  extends over all integer partitions $k\in\mathcal{P}(2\omega;\sigma_{ij})$ of the even integer ${2\omega=\sum_{i=1}^n \sigma_i}$ into the $n(n+1)/2$ integers $\sigma_{ij}$ compatible with the $n$ linear constraint equations $\sigma_i=2\sigma_{ij}+\Sigma_i$ with $\Sigma_{i}:=\sum_{ j\neq i=1}^{n}\sigma_{ij}$. The coefficient $C_{k}$ corresponds to the combinatorial weight of the $k$th partition and was derived explicitly in \cite{Steinwachs2019},
\begin{align}
C_k=\frac{\prod_{j=1}^{n}\left[\Sigma_j!
	\left(
	\begin{array}{c}
	2\sigma_{jj}+\Sigma_j
	\\
	\Sigma_j
	\end{array}
	\right)
	(2\sigma_{jj}-1)!!\right]}{\prod_{1\leq i\leq j\leq n}\sigma_{ij}!},\label{Cweight}
\end{align}
Finally, $P_{\omega}(d)$ is a polynomial of order $\omega$ in the dimension $d$ defined by
\begin{align}
P_\omega(d):=\prod_{i=1}^{\omega}\left[d+2(i-1)\right]=\frac{2^\omega\Gamma(\omega+d/2)}{\Gamma(d/2)}\label{DimPol}
\end{align}
and we have absorbed the pole in dimension together with a numerical factor in the definition
\begin{align}
\Lambda:=\frac{1}{(4\pi)^2\varepsilon}
\end{align}
Summing over all contributions of all logarithmically divergent integrals \eqref{DivIntFin} arising from the propagator expansion \eqref{PropEx}, yields the divergent part of the integrals $I_k$ defined in \eqref{FeynIntn}. Summation of all integrals $I_k$, we obtain the divergent part of the original integral $I_X=\sum_kI_{k}$ corresponding to the $1$PI diagram $X=2a,\ldots,5e$.
 
Before we discuss the application of this algorithm to the Galileon model, we first comment on the use of dimensional regularization in connection with identities involving the Levi-Civita symbol. Some care is needed when replacing the dimension from four to $d=4-2\varepsilon$ in the expansion of the 
Levi-Civita symbols in \eqref{V3}-\eqref{V5}, as the following dimensional dependent identity is valid only for $d=n$,
\begin{align}
\epsilon_{\mu_1\mu_2\ldots\mu_n}\epsilon^{\nu_1\nu_2\ldots\nu_n}=\left|
\begin{array}{ccc}
\delta_{\mu_1}^{\nu_1}&\ldots&\delta_{\mu_1}^{\nu_n}\\
\vdots&{}&\vdots\\
\delta_{\mu_n}^{\nu_1}&\ldots&\delta_{\mu_n}^{\nu_n}
\end{array}
\right|.\label{DimDepId}
\end{align}
Despite the mismatch for $n=4$ in the Galileon action \eqref{Gact} when using dimensional regularization in $d=4-2\varepsilon\neq4$ dimensions, when the identity \eqref{DimDepId} is formally expanded around $\varepsilon=0$, the correction terms are $\mathcal{O}\left(\varepsilon\right)$. Therefore, for the calculation of the on-loop divergences, it is legitimate to use the identity \eqref{DimDepId} for $n=d=4$, as any $\mathcal{O}\left(\varepsilon\right)$ contribution which combines with the leading UV pole $1/\varepsilon$ would already be finite or vanish in the limit $\varepsilon\to0$.

\section{Representation of the result and crosschecks}
\label{SecRepResCheck}
Since the divergent part of the $n$-point correlation function  
\begin{align}
\langle x_1,x_2,\ldots,x_n \rangle^{\mathrm{div}}_{1}:=\left.\frac{\delta^n\Gamma^{\mathrm{div}}_{1}}{\delta\pi(x_1)\ldots\delta\pi(x_n)}\right|_{\pi=0},\label{CorrelationFunction}
\end{align}
is totally symmetric under exchange of the $\pi_i$, $i=1,\ldots,n$, the Feynman integrals $I^{\mathrm{div}}_X$ for the topologies $X=2a,\ldots,5e$ must be totally symmetric functions of the external momenta $k_{i}^{\mu}$, $i=1,\ldots,n$. In order to obtain a representation for $I^{\mathrm{div}}_X$ in which this symmetry is manifest, we could proceed as follows: For each integral $I^{\mathrm{div}}_X$ we choose a particular kinematic parametrization in terms of the $q_i^{\mu}$, for which momentum conservation $\sum_{i=1}^n q_{i}^{\mu}=0$ has been implemented. The final result for the one-loop divergences is expressed in terms of the external momenta $k_{i}^{\mu}$ by inverting relation \eqref{exq} and subsequently symmetrized among the $n$ external momenta with unit weight $1/n!$. Finally, the result for each diagram is multiplied by the multiplicity $M$ and the symmetry factor $S$.  

In general, a scalar $n$-point integral depends on $3n-10$ independent invariants because there are $n$ external momenta $p_{i}^{\mu}$, each having $4$ independent components in $d=4$ dimensions minus $n$ constraint equations from the on-shell relation $p_i^2=m_i^2$ and $10$ constraint equations from the invariance under Poincar\'e transformations, since the Poincar\'e group in $d=4$ has $10$ generators. In contrast, there are $n(n+1)/2$ possible quadratic scalar invariants $G_{ij}$, defined in \eqref{Gram}, which can be constructed from the scalar products among the $n$ external momenta. Momentum conservation $\sum_{i=1}^n k_{i}^{\mu}=0$ allows to arbitrarily pick one external momentum $k_{i}^{\mu}$ and express it in terms of the $n-1$ remaining external momenta. Therefore, the number of independent off-shell invariants is $n(n-1)/2$. Imposing the $n$ on-shell conditions $p_i^2=m_i^2$ reduces the number of independent on-shell invariants further to $n(n-3)/2$, which can be chosen as a subset of the $n(n-1)/2$ Mandelstam variables
\begin{align}
s_{ij}:=(k_{i}+k_{j})^2,\qquad i\neq j.\label{Mandelstam}
\end{align}
In particular, this implies that the on-shell result for the $3$-point function vanish trivially due to kinematics. Moreover, for $n=4$ and $n=5$, the equality $3n-10=n(n-3)/2$ holds. For $n\geq6$ there is a redundancy in the $n(n-3)/2$ bilinear invariants due to the fact that the $n-1\geq 5$ momenta are linearly dependent in $d=4$ dimensions. The additional quadratic constraint equations can be easily derived \cite{Asribekov1962}. 

For a compact representation of the results, it might however be beneficial to reintroduce a redundancy among the invariants and express the final result in terms of an enlarged set of variables.
For example, in case of the $4$-point function the adjacent Mandelstam variables $s_{12}$, $s_{23}$ and $s_{31}$, which close under cyclic permutations, are a convenient set of redundant on-shell variables, for which the result can be compactly expressed in terms of power-summed elementary symmetric polynomials. The off-shell results become increasingly complicated and the number of terms grow at least by an order of magnitude for each additional external leg. The number of terms for the $2$-point off-shell result is $\mathcal{O}(1)$, for the $3$-point $\mathcal{O}(10)$, for the $4$-point $\mathcal{O}(10^2)$ and for the $5$-point even $\mathcal{O}(10^4)$. We therefore refrain from explicitly presenting the results in the main text or in an appendix but instead append an ancillary file in which the results for the divergent part of the off-shell $n$-point functions for $n=1,\ldots,5$ are stored in an ancillary file.    

As explained in Sec. \ref{algorithm}, we extract the off-shell divergences of the diagrams ${2a,\ldots,5e}$ by the general combinatorial algorithm introduced in \cite{Steinwachs2019}. There are however certain diagrams, for which it is obvious that the associated integrals identically vanish. It is e.g. clear that the tadpole diagrams $I^{\mathrm{div}}_{2b}$ and $I^{\mathrm{div}}_{3c}$ are scale-free and therefore trivially vanish when using dimensional regularization. Below, we provide the result of the off-shell divergences for all the $1$PI one-loop $2$-point and $3$-point diagrams as well as the on-shell divergences for the $1$PI one-loop $4$-point diagrams and compare them with previous results in the literature.

\subsection{Divergent off-shell one-loop contributions of $1$PI $2$-point diagrams}
The divergent off-shell contributions to the $1$PI one-loop diagrams $2a,\ldots,2b$ are
\begin{align}
I^{\mathrm{div}}_{2a}={}&\frac{9}{2}\frac{\Lambda c_3^2}{ M^6 }\label{2pf} \left(k_{1}^2\right)^4,\\
I^{\mathrm{div}}_{2b}={}&0.
\end{align}
The result \eqref{2pf} perfectly agrees with the result obtained in eq. (28) of \cite{PaulaNetto2012}.

\subsection{Divergent off-shell one-loop contributions of $1$PI $3$-point diagrams}
The divergent off-shell contributions to the $1$PI one-loop diagrams $3a,\ldots,3c$ are
\begin{align}
I^{\mathrm{div}}_{3a}={}\frac{9}{4}\frac{\Lambda c_3^3}{M^9}&\Big\{3  (k_1^2)^5-  (k_1^2)^4 k_{22}-2  (k_1^2)^3 (k_2^2)^2-2 (k_1^2)^2 (k_2^2)^3-k_{11} (k_2^2)^4+3 (k_2^2)^5\nonumber\\
&-s_{12} \left[(k_1^2)^4-4 (k_1^2)^3 k_{22}-14 (k_1^2)^2 (k_2^2)^2-4 k_{11} 
(k_2^2)^3+(k_2^2)^4\right]\nonumber\\
&- 2 s_{12}^2 \left[(k_1^2)^3-7 (k_1^2)^2 k_{22}-7 k_{11} (k_2^2)^2+(k_2^2)^3\right]-2  s_{12}^3 (k_{11}-k_{22})^2\nonumber\\
& - s_{12}^4 (k_{11}+k_{22})+3  s_{12}^5\Big\},\label{I3a}\\
I^{\mathrm{div}}_{3b}={}\frac{3}{8}\frac{\Lambda c_3c_4}{M^9}&\Big\{6 (k_1^2)^5-7 (k_1^2)^4 k_{22}+2   (k_1^2)^3 (k_2^2)^2+2  (k_1^2)^2 (k_2^2)^3-7 k_{11} (k_2^2)^4+6   (k_2^2)^5\nonumber\\
&-s_{12} \left[7 (k_1^2)^4+8 (k_1^2)^3 k_{22}+4 (k_1^2)^2 (k_2^2)^2+8 k_{11} (k_2^2)^3+7 (k_2^2)^4\right]\nonumber\\
&+2  (s_{12})^2 (k_{11}-k_{22})^2 (k_{11}+k_{22}) +2 s_{12}^3 \left[(k_1^2)^2+(k_2^2)^2\right]\Big\}\label{I3b},\\
I^{\mathrm{div}}_{3c}={}0.\phantom{\frac{9}{4}\frac{c_3^3}{M^9\varepsilon}}&
\end{align}
As required for kinematic reasons, the on-shell reduction $k_1^2=k_2^2=k_3^2=0$ of \eqref{I3a} and \eqref{I3b} leads to $I_{3a}^{\mathrm{div}}=I_{3b}^{\mathrm{div}}=0$, since $s_{12}=(k_1+k_2)^2=k_3^2=0$.

\subsection{Divergent on-shell one-loop contributions of $1$PI $4$-point diagrams}
Due to the aforementioned counting of $n(n-3)/2$ independent quadratic invariants for an $n$-point diagram, it is clear that the first non-trivial result for the on-shell one-loop divergences starts with the $4$-point function. We therefore provide the on-shell reduction of the off-shell result for the divergent one-loop $4$-point contributions. The on-shell result for the $4$-point function only depends on two independent Mandelstam variables, say $s_{12}$ and $s_{23}$. However, a more compact and symmetric representation is obtained by reintroducing the redundant invariant $s_{13}$ and by expressing the result in terms of a power-summed symmetric polynomial of the three Mandelstam variables $s_{12}$, $s_{13}$ and $s_{23}$, for which the divergent on-shell contributions to the $1$PI one-loop diagrams $4a,\ldots,4d$ take the compact form
\begin{align}
I^{\mathrm{div, on-shell}}_{4a}={}&\frac{243}{80}\frac{\Lambda c_3^4}{ M^{12}}\left(s_{12}^2+s_{23}^2+s_{31}^2\right)^3,\label{4adiv}\\
I^{\mathrm{div, on-shell}}_{4b}={}&\frac{9}{20}\frac{\Lambda c_3^2c_4}{ M^{12}}\left[20\left(s_{12}^6+s_{23}^6+s_{31}^6\right)-3\left(s_{12}^2+s_{23}^2+s_{31}^2\right)^3\right],\\
I^{\mathrm{div, on-shell}}_{4c}={}&\frac{3}{20}\frac{\Lambda c_4^2}{ M^{12}}\left(s_{12}^2+s_{23}^2+s_{31}^2\right)^3,\\
I^{\mathrm{div, on-shell}}_{4d}={}&0\label{4ddiv}.
\end{align}
The expressions \eqref{4adiv}-\eqref{4ddiv} coincide with those of the $1$PI diagrams presented in Table $1$ of \cite{Kampf2014} upon identification of $\Lambda\to-2\tilde{\Lambda}$, where we denote by $\tilde{\Lambda}$ the rescaled pole in dimension as defined in eq. (6.221) of \cite{Kampf2014}.

\section{Renormalization, effective field theory and resummation}
\label{SecRenEff}
In this section, we discuss the structure of the loop induced operators in the Galileon effective field theory approach and comment on various resummation schemes.
As discussed in \cite{Brouzakis2014,Brouzakis2014a}, the effective action has the general schematic structure 
\begin{align}
\Gamma_{1}=\int\mathrm{d}^4x\sum_k\left[M^4+M^2\partial^2+\partial^4\log\left(\frac{\partial^2}{M^2}\right)\right]\left(\frac{\partial^2\pi}{M^3}\right)^k,\label{EffActStruc}
\end{align}
where we have identified $M$ as the cutoff scale and suppressed the index structure.
In dimensional regularization only the last term survives and upon identification of the logarithm with the pole in dimension corresponds to the UV divergent part.
By inspection of \eqref{EffActStruc}, it is clear that there are two parameters which control the hierarchy among different operators in the Galileon effective field theory expansion (suppressing again the index structure)
\begin{align}
\sigma_{\mathrm{\partial}}:=\frac{\partial^2}{M^2},\qquad \sigma_{\mathrm{\partial^2\pi}}:=\frac{\partial^2\pi}{M^3}\label{sig}.
\end{align} 
The first parameter measures the expansion in derivatives, which for the Galileon action \eqref{Gact} is related to an expansion in the number of loops as the loop induced operators come with a higher number of derivatives per field than the tree-level operators \eqref{L0}-\eqref{L4}. The second parameter measures the degree of non-linearity, which is related to an expansion in the number of external legs. The table shows counter term operators of the form
\begin{align}
\mathcal{L}_{\ell,n}^{\mathrm{div}}.
\end{align}
These operators are classified by two indices.
The first subindex counts the loop order at which the operator is induced (given the tree--level operators $\mathcal{L}_{0,n}^{\mathrm{tree}}$) and is related to powers of $\sigma_{\partial}$. The second subindex counts the number of fields $\pi$ 
and is related to powers of $\sigma_\mathrm{\partial^2\pi}$.
\begin{table}[h!]
\begin{center}
\begin{tabular}{c|cccccc}
$\sigma_{\partial}/\sigma_{\partial^2\pi}$&0&1&2&3&\ldots\\
\hline
&&&&&&\\[-1em] 
0&$\mathcal{L}_{0,2}^{\mathrm{tree}}$&$\mathcal{L}_{0,3}^{\mathrm{tree}}$&$\mathcal{L}_{0,4}^{\mathrm{tree}}$&$\mathcal{L}_{0,5}^{\mathrm{tree}}$&\ldots\\
&&&&&&\\[-1em] 
3&$\mathcal{L}_{1,2}^{\mathrm{div}}$&$\mathcal{L}_{1,3}^{\mathrm{div}}$&$\mathcal{L}_{1,4}^{\mathrm{div}}$&$\mathcal{L}_{1,5}^{\mathrm{div}}$&\ldots\\
&&&&&&\\[-1em] 
6&$\mathcal{L}_{2,2}^{\mathrm{div}}$&$\mathcal{L}_{2,3}^{\mathrm{div}}$&$\mathcal{L}_{2,4}^{\mathrm{div}}$&$\mathcal{L}_{2,5}^{\mathrm{div}}$&\ldots\\
\vdots&\vdots&\vdots&\vdots&\vdots&$\ddots$
\end{tabular}
\end{center} 
\caption{Structure of higher dimensional operators in an effective field theory expansion of the Galileon, ordered by different powers of the dimensionless parameters \eqref{sig}.}
\label{Table1}
\end{table}

Given the ordering scheme presented in Table \ref{Table1}, two different ways of resummations these operators suggest themselves. First, a resummation of terms with arbitrary derivatives $\partial$ but a fixed number of fields $\pi$ would lead to structures such as
\begin{align}
\mathcal{L}_{\infty,2}^{\mathrm{div}}:=\sum_{k=1}^{\infty}\mathcal{L}_{k,2}^{\mathrm{div}}=-\frac{1}{2}\pi\, f\left(\frac{\partial^2}{M^2}\right)\partial^2\pi.\label{Resummed}
\end{align} 
The resummation in \eqref{Resummed} would incorporating all operators bilinear in $\pi$ and lead to (possibly non-local) form factors $f(\partial^2/M^2)$, which upon expansion in powers of $\partial/M^2$,
\begin{align} f\left(\frac{\partial^2}{M^2}\right)=1+c_1\left(\frac{\partial^2}{M^2}\right)^3+\ldots 
\end{align}
must recover the terms in the first column in Table \ref{Table1} with the appropriate numerical coefficients $c_1$,$\ldots$
For higher powers of the field $\pi$, the resummation of different index contractions must be taken into account and at each given order with a fixed number of fields one has to find a finite basis of operators, each with a different form factor. In the context of gravity and the heat-kernel technique, such a resummation has been discussed in \cite{Barvinsky1987,Barvinsky1990}.

Instead of resumming terms with a fixed number of fields and an arbitrary number of derivatives, it would also be interesting to consider the opposite case, which would correspond to resumming terms with a fixed number of derivatives per field but an arbitrary number of fields. This resummation would incorporate operators with an arbitrary power of $\sigma_{\partial^2\pi}$ at a fixed order of $\sigma_{\partial^2}$. In fact, such a resummation shares many similarities to General Relativity, where an arbitrary number of metric perturbations with two derivatives are geometrically resummed in the Ricci scalar. In fact, starting with the linear equations of motion for a massless spin-two field propagating on flat spacetime and taking into account contributions to the energy momentum tensor of the gravitational field in a self-consistent way, requires to iteratively introduce particular self-interaction terms, which upon resummation lead to the full non-linear theory of General Relativity, see e.g. \cite{Deser1970}.

Another interesting resummation of tree-level operators in the context of scalar effective field theories has been carried out in \cite{Cheung2015,Cheung2015a,Cheung2017}. There, the behavior of tree-level scattering amplitudes under soft and colinear limits were used to bootstrap various scalar field theories. In case of the special Galileon theory, it turns out that the constraints are so strong that they fix all coupling constants of higher dimensional operators in terms of the coupling constants of lower dimensional operators. Resummation of the whole series then results in a square-root DBI-type action for the scalar field. 

In an upcoming work \cite{Heisenberg2019}, we investigate whether a resummation of all divergent one-loop $n$-point functions of the scalar Galileon can be found in the form of geometrically defined effective action, which would then serve as a generating functional for all $n$-point counterterms.   

\section{Conclusions}
\label{SecCon}
In this article we have investigated the renormalization structure of the scalar Galileon model in flat spacetime. We performed a Feynman diagrammatic calculation of the UV divergent one-loop contributions to the off-shell $2$-point, $3$-point, $4$-point and $5$-point functions. This generalizes the previously known results for the off-shell $2$-point function \cite{PaulaNetto2012} and the on-shell $4$-point function \cite{Kampf2014}. The actual calculation has been performed with the recently suggested combinatorial algorithm designed to efficiently extract the one-loop divergences in higher derivative field theories \cite{Steinwachs2019}. In view of the large number of terms, which involve $\mathcal{O}(10^2)$ terms for the $4$-point function and $\mathcal{O}(10^4)$ terms for the $5$-point function, we have provided the explicit results for the UV divergences in an ancillary file. We have checked our results with previous calculations of the off-shell $2$-point and on-shell $4$-point functions and found perfect agreement.

We discussed the counterterm structure and the hierarchy of the loop-induced Galileon operators in an effective field theory framework. We identified two natural expansion parameters \eqref{sig}, which control the effective field theory expansion and are associated with the derivative order (related to the loop order) and the non-linearity of the theory, respectively.
We discussed two different types of resummation schemes. The first corresponds to a resummation of terms with arbitrary orders of $\sigma_{\partial}$, i.e. to a resummation of terms with a fixed number of fields $\pi$ but an arbitrary number of derivatives $\partial$. The resummed operators might be compactly written in terms of form factors as e.g. in the context of the covariant perturbation theory discussed in \cite{Barvinsky1987,Barvinsky1990}. The second, corresponds to a resummation of terms with arbitrary powers of $\sigma_{\partial^2\pi}$, i.e. to operators with a fixed number of derivatives $\partial^2\pi$ per field but an arbitrary number of fields $\pi$. Such a resummation, is very similar to the geometrical resummation of the self-interaction terms, starting from the linear field equations for a massless spin-two particle propagating in flat spacetime and recovering the full non-linear theory of General Relativity \cite{Deser1970}. 
In an upcoming work, we investigate a generalization of our result based on a geometric resummation of the divergent one-loop Galileon $n$-point functions \cite{Heisenberg2019}.

\acknowledgments
L.H. specially thanks Karol Kampf and Jiri Novotny for useful skype calls. C.S. thanks Samuel Abreu, Stefan Dittmaier, Paul McFadden, Michael Ruf and Mao Zeng for interesting discussions. LH is supported by funding from the European Research Council (ERC) under the European Unions Horizon 2020 research and innovation programme grant agreement No 801781 and by the Swiss National Science Foundation grant 179740.

\bibliographystyle{JHEP}
\bibliography{FeynmanGalileon}{}
\end{document}